\documentclass[12pt,english,aps,manuscript]{article}
\usepackage[T1]{fontenc}
\usepackage[latin9]{inputenc}
\usepackage{geometry}
\usepackage{amsmath}
\usepackage{amssymb}
\usepackage[numbers]{natbib}
\usepackage{babel}
\begin{document}
\begin{center}
\textbf{\Large{}Notes on Quantum Effective Actions}{\Large\par}
\par\end{center}

\begin{center}
\vspace{0.3cm}
\par\end{center}

\begin{center}
{\large{}S. P. de Alwis$^{\dagger}$ }{\large\par}
\par\end{center}

\begin{center}
Physics Department, University of Colorado, \\
 Boulder, CO 80309 USA 
\par\end{center}

\begin{center}
\vspace{0.3cm}
 
\par\end{center}

\begin{center}
\textbf{Abstract} 
\par\end{center}

\begin{center}
\vspace{0.3cm}
\par\end{center}

\smallskip{}
\vspace{0.3cm}
 We first note that, at least in perturbation theory, there is a well-defined
(subject to regularization) Lorentzian definition of the quantum effective
action in both flat and curved space including (perturbative) gravity.
The advantage of the latter is that we do not need to deal with the
conformal factor problems of Euclidean quantum gravity. We then make
some remarks on the Euclidean version (in flat space) and convexity
and resolve a puzzle that highlights the importance of keeping the
initial and final states in the functional integral. Next we discuss
the gauge invariant effective action of Vilkovisky and DeWitt and
show its gauge fixing independence. We conclude with the expression
for the Wilsonian effective action in this framework.
\begin{center}
\vspace{0.3cm}
 
\par\end{center}

\vfill{}

$^{\dagger}$ dealwiss@colorado.edu

\eject

\section{Quantum Effective Action in Lorenzian QFT}

\subsection{Flat space}

The path integral was introduced by Feynman as a formal solution to
the Schroedinger equation, not the diffusion equation, and is a representation
of the matrix elements of the unitary operator $e^{-i\hat{H}t}$.
In QFT this is represented by the Lorenzian path integral. The generator
of connected correlation functions is given by the (schematic) formula\footnote{We will be using the condensed DeWitt notation in these notes. So
$j.\phi=\int d^{4}x\sqrt{|g|}j(x)\phi(x)$ , etc.}
\begin{equation}
e^{iW[j]}=<O|Te^{ij.\hat{\phi}}|I>=\int[d\phi]e^{iS[\phi]+ij.\phi}\label{eq:Wj-1}
\end{equation}
We have omitted the initial and final wave fucntions on the RHS of
the last equation for simplicity of presentation. \footnote{Later we will discuss the importance of keeping them in order to avoid
some paradoxes. Since in the following we are only going to show how
a well-defined perturbation series arises (with a view to extending
it to quantum gravity), we can safely omit writing them explicitly
and just invoke them when needed} \footnote{For a rescent discussion of Lorenzian arguments for the effective
action in curved space at the one-loop level see \citep{Baldazzi:2019kim}. }. 

Defining the quantum effective action in this Lorenzian framework
as usual by $\Gamma[\phi_{c}]=W[j]-j.\phi_{c},\,\phi_{c}=\delta W/\delta j\implies j=-\delta\Gamma/\delta\phi_{c}$
we have the recursive expression
\begin{align}
e^{i\Gamma[\phi_{c}]} & =\int[d\phi]e^{iS[\phi]+ij[\phi_{c}].(\phi-\phi_{c})}|_{j=-\delta\Gamma/\delta\phi_{c}}\nonumber \\
 & =\int[d\phi']e^{iS[\phi_{c}+\phi']+ij[\phi_{c}].(\phi')}|_{j=-\delta\Gamma/\delta\phi_{c}}\nonumber \\
 & =\int[d\phi']e^{i\left[S[\phi_{c}]+\frac{1}{2}\phi'.K[\phi_{c}].\phi'+S_{I}\left[\phi_{c},\phi'\right]+\bar{j}[\phi_{c}].\phi'\right]}|_{\bar{j}=-\delta\Gamma/\delta\phi_{c}+\delta S/\delta\phi_{c}}.\label{eq:PIforGamma}
\end{align}
Here the background field interaction action $S_{I}[\phi_{c},\phi']$
is at least cubic in the 'quantum' field $\phi'$ and the background
field dependent kinetic matrix is 
\begin{equation}
K[\phi_{c};x,y]=\frac{\delta^{2}S[\phi_{c}]}{\delta\phi_{c}(x)\delta\phi_{c}\left(y\right)}.\label{eq:K}
\end{equation}
Note that $K$ is a matrix in field space as well as in space-time.
The term $S_{I}[\phi']$ may be taken outside the functional integral
by using $\delta/i\delta j\leftrightarrow\phi'$, and doing the resulting
Gaussian integral\footnote{We will discuss such integrals involving exponentials with an imaginary
argument for hyperboic differential operators presently.} we have,
\begin{equation}
e^{i\Gamma[\phi_{c}]}=e^{iS[\phi_{c}]}e^{-\frac{1}{2}{\rm Tr\ln K\left[\phi_{c}\right]}}e^{iS_{I}\left[\phi_{c},\frac{\delta}{i\delta j}\right]}e^{-\frac{i}{2}\bar{j}.K[\phi_{c}]^{-1}.\bar{j}}|_{\bar{j}=-\delta\Gamma/\delta\phi_{c}+\delta S/\delta\phi_{c}}.\label{eq:ExpGamma_expansion}
\end{equation}
 Using now a standard identity (see for example \citep{Coleman:1985rnk})
we have a (formal) expression for the quantum effective action,
\begin{align}
\Gamma[\phi_{c}] & =S[\phi_{c}]+\frac{i}{2}{\rm Tr\ln K[\phi_{c}]}\nonumber \\
 & +i^{-1}\ln\left\{ \exp\left(-\frac{i}{2}\frac{\delta}{i\delta\phi'}.K^{-1}.\frac{\delta}{i\delta\phi'}\right)\exp\left(iS_{I}\left[\phi_{c},\phi'\right]+\left(-\delta\Gamma/\delta\phi_{c}+\delta S/\delta\phi_{c}\right).\phi'\right)\right\} _{\phi'=0}.\label{eq:Gammaexpansion}
\end{align}
However we know that $\Gamma$ only involves 1PI graphs. Hence the
linear term inside the log can be ignored and replaced with an instruction
to compute just 1PI graphs. Note that the first line on the RHS of
this equation gives the zero loop (classical action) and the one loop
contribution to $\Gamma$ while the second line gives all higher loop
contributions.

The above expression is formal and we need to regularize the divergences
that arise from propagators at concident space time points. For our
purposes a convenient method is to introduce the so-called heat kernel
representation \citep{DeWitt:1964mxt}\footnote{For a review see for example \citep{Vassilevich:2003xt}.},
\begin{equation}
K^{-1}\left[\phi_{c};x,y\right]=-i\int_{0}^{\infty}ds<x|e^{is\left(\hat{K}[\phi_{c}]+i\epsilon\right)}|y>,\label{eq:Kinverse}
\end{equation}
where we have incorporated the $i\epsilon$ term coming from the wave
functions \citep{Weinberg:1996kr} which ensures that the contribution
to the integral from the upper limit vanishes. We have introduced
the quantum mechanics notation with position operator $\hat{x}$ and
eignestates defined by $\hat{x}^{\mu}|x>=x^{\mu}|x>$ with normalization
$<x|y>=\delta^{4}(x-y)$, momentum operator $\hat{p}_{\mu}$ and eigenstates
$\hat{p}_{\mu}|p>=p_{\mu}|p>$, completeness relation $\int d^{4}p|p><p|=\hat{I}$
the unit operator, and $<x|p>=\frac{e^{ip.x}}{(2\pi\}^{2}}$. The
(Lorenzian) heat kernel $H$ is defined \citep{DeWitt:1964mxt} as
the solution of the Schroedinger eqn 
\begin{equation}
i\frac{d}{ds}H(s|x,y)=-\hat{K}H(s|x,y),\label{eq:Heateqn}
\end{equation}
and hence has the representation
\begin{equation}
H(s|x,y)=<x|e^{is\left(\hat{K}[\phi_{c}]+i\epsilon\right)}|y>.\label{eq:Hflat}
\end{equation}
Let us first consider the $\phi_{c}={\rm constant}$ case - this is
sufficient for calculating the effective potential. So we get\footnote{The $p$ integrals below are evaluated by using the formula $\int_{-\infty}^{\infty}e^{ix^{2}}dx=e^{i\pi/4}\sqrt{\pi}$
which can be obtained from the usual Gaussian integral by applying
Cauchy's theorem to a contuor integral over a wedge with one side
being the real line and the other a straight line passing through
the origin and oriented at an angle $\pi/4$ to it.} (by inserting a complete set of $p$ eigenstates in \eqref{eq:Hflat})
\begin{equation}
H(s|x,y)=e^{-isV''(\phi_{c})}\int\frac{d^{4}p}{(2\pi)^{4}}e^{-isp^{2}+i\left(x-y\right)^{\mu}p_{\mu}}=-i\frac{1}{16\pi^{2}s^{2}}e^{-isV^{''}}e^{\frac{i}{4s}\left(x-y\right)^{2}}.\label{eq:Hflat2}
\end{equation}
The background field propagator is then 
\begin{equation}
K^{-1}\left[\phi_{c};x,y\right]=-\int_{0}^{\infty}\frac{ds}{16\pi^{2}s^{2}}e^{-isV''(\phi_{c})}e^{\frac{i}{4s}\left(x-y\right)^{2}}\label{eq:Kinverse1}
\end{equation}
Now we apply Cauchy's theorem to the following contour. This consists
of the real line $\delta\rightarrow R$ ($R\gg\delta>0)$, quarter
circles in the south-east quadrant of radius $R$ and $\delta$, and
the imaginary segment $-Ri\rightarrow-i\delta$. Taking the limit
$\delta\rightarrow0,\,R\rightarrow\infty$, we see that we can rewrite
this integral as (effectively writing $s=-i\tau$ and provided that
$\left(x-y\right)^{2}>0$ i.e. space-like, and $V,''>0$ i.e. the
fields are non-tachyonic),
\begin{equation}
K^{-1}\left[\phi_{c};x,y\right]=-i\int_{0}^{\infty}\frac{d\tau}{16\pi^{2}\tau^{2}}e^{-\tau V''(\phi_{c})}e^{-\frac{1}{4\tau}\left(x-y\right)^{2}}.\label{eq:Kinverse2}
\end{equation}
Similarly we can get a well-defined expression for the log of $K$,
\begin{align}
\ln K[\phi_{c}] & =-\int_{0}^{\infty}\frac{ds}{s}<x|e^{is\hat{K}[\phi_{c}]}|y>\nonumber \\
 & =i\int_{0}^{\infty}\frac{d\tau}{16\pi^{2}\tau^{3}}e^{-\tau V''(\phi_{c})}e^{-\frac{1}{4\tau}\left(x-y\right)^{2}}.\label{eq:logK}
\end{align}
These two expressions are well-defined as long as the interval is
space-like separated but diverge on the light-cone $(x-y)^{2}=0$.
To define them everywhere we need to introduce a regulator. A convenient
way of doing this is to cutoff the integral at its lower divergent
end. Thus we define the regulated propagator and its log by,
\begin{align}
K_{\Lambda}^{-1}\left[\phi_{c};x,y\right] & =-i\int_{1/\Lambda^{2}}^{\infty}\frac{d\tau}{16\pi^{2}\tau^{2}}e^{-\tau V''(\phi_{c})}e^{-\frac{1}{4\tau}\left(x-y\right)^{2}}\label{eq:Kinverse_constphiReg}\\
\ln K_{\Lambda}[\phi_{c}] & =i\int_{1/\Lambda^{2}}^{\infty}\frac{d\tau}{16\pi^{2}\tau^{3}}e^{-\tau V''(\phi_{c})}e^{-\frac{1}{4\tau}\left(x-y\right)^{2}}\label{eq:logKreg}
\end{align}

Recall that we've used a condensed notation and the above equations
are actually realtions between matrices - both in field space as well
as in space-time. A regulated expression for the one-loop effective
potential is then obtained by taking the trace over all indices of
the last equation.
\begin{equation}
\Gamma_{\Lambda}^{(1)}[\phi_{c}]=i\frac{1}{2}{\rm Tr\ln K_{\Lambda}[\phi_{c}]=-\frac{1}{2}\int d^{4}x}\int_{1/\Lambda^{2}}^{\infty}\frac{d\tau}{16\pi^{2}\tau^{3}}{\rm tr}e^{-\tau V''(\phi_{c})}.\label{eq:Gamma1constphiReg}
\end{equation}
Going beyond constant $\phi_{c}$, the complete regularized perturbation
series for $\Gamma$ is then obtained by rewriting \eqref{eq:Gammaexpansion}
as 
\[
e^{i\Gamma[\phi_{c}]}=e^{iS_{\Lambda}[\phi_{c}]}e^{-\frac{1}{2}{\rm Tr\ln K_{\Lambda}\left[\phi_{c}\right]}}e^{iS_{I,\Lambda}\left[\phi_{c},\frac{\delta}{i\delta j}\right]}e^{-\frac{i}{2}\bar{j}.K_{\Lambda}[\phi_{c}]^{-1}.\bar{j}}|_{\bar{j}=-\delta\Gamma/\delta\phi_{c}+\delta S_{\Lambda}/\delta\phi_{c}},
\]
or 
\begin{align*}
\Gamma[\phi_{c}] & =S_{\Lambda}[\phi_{c}]+\frac{i}{2}{\rm Tr\ln K_{\Lambda}[\phi_{c}]}\\
 & +i^{-1}\ln\left\{ \exp\left(-\frac{i}{2}\frac{\delta}{i\delta\phi'}.K_{\Lambda}^{-1}[\phi_{c}].\frac{\delta}{i\delta\phi'}\right)\exp\left(iS_{I,\Lambda}\left[\phi_{c},\phi'\right]+\left(-\delta\Gamma/\delta\phi_{c}+\delta S_{\Lambda}/\delta\phi_{c}\right).\phi'\right)\right\} _{\phi'=0}.
\end{align*}
Here the regularized expressions for the propagator and one-loop effective
action are given by (we've used bold face letters below to emphaize
the fact that they are matrices in field space),
\begin{align}
{\bf K}_{\Lambda}^{-1}\left[\phi_{c};x,y\right] & =-i\int_{1/\Lambda^{2}}^{\infty}ds<x|e^{-is\left(\hat{{\bf K}}[\phi_{c}]+i\epsilon\right)}|y>,\label{eq:KinverseReg}\\
 & =-i\int_{1/\Lambda^{2}}^{\infty}\frac{d\tau}{16\pi^{2}\tau^{2}}e^{-\frac{1}{4\tau}\left(x-y\right)^{2}}\sum_{n=0}^{\infty}{\bf a}_{2n}(x,y)\tau^{n},\label{eq:KinverseReg2}
\end{align}
and 
\begin{align}
\Gamma_{\Lambda}^{(1)}[\phi_{c}] & =i\frac{1}{2}{\rm Tr\ln{\bf K}_{\Lambda}[\phi_{c}]=-\frac{1}{2}i\int d^{4}x}\int_{1/\Lambda^{2}}^{\infty}\frac{ds}{s}{\rm tr}<x|e^{-is\left(\hat{{\bf K}}[\phi_{c}]+i\epsilon\right)}|x>\nonumber \\
 & =-\frac{1}{2}\int d^{4}x\int_{1/\Lambda^{2}}^{\infty}\frac{d\tau}{16\pi^{2}\tau^{3}}{\rm tr}\sum_{n=0}^{\infty}{\bf a}_{2n}(x,x)\tau^{n}.\label{eq:Gamma1Reg}
\end{align}
In the second line of the two equations above we've done the same
analytic continuation as in the constant $\phi_{c}$ case - the only
difference is that we have a infinite series the so-called heat kernel
expansion \citep{DeWitt:1964mxt}. The heat kernel expansion is only
valid for short proper times (i.e. $\tau$) compared to time scales
in the theory -so the $\tau$ integrals in the above eqns should be
cutoff at some scale $1/\Lambda'^{2}>1/\Lambda^{2}$. This would imply
that what we are defining is the Wilsonian effective action at the
scale $\Lambda'$ rather than the 1PI action. The former is given
as an infinite series of local operators whilst one would expect the
1PI action to also have non-local terms. This is the general situation
since one necessarily has massless fields in a realistic theory -
gauge and gravitational fields. On the other hand if the lowest mass
in the theory is non-zero one has an explicit infra-red cutoff (i.e.
an explicit factor $e^{-m^{2}{\bf I}\tau}$ in the $\tau$ integrand
with corresponding modifications in the heat kernel coefficients ${\bf a}_{n}$,
and it is then possible to get a one-loop expression for the 1PI action,
as an infinite series of local terms.

Let us discuss the latter situation with just a massive scalar field.
In this case ${\bf K}=\nabla^{2}-V''$ So we need to compute
\begin{equation}
\Gamma_{\Lambda}^{(1)}[\phi_{c}]=-\frac{1}{2}\int d^{4}x\int_{1/\Lambda^{2}}^{\infty}\frac{d\tau}{16\pi^{2}\tau^{3}}{\rm tr}e^{-\tau m^{2}{\bf I}}\sum_{n=0}^{\infty}\tilde{{\bf a}}_{2n}(x,x)\tau^{n},\label{eq:Gamma1scalar}
\end{equation}
 and the first few heat kernel coefficients (with $V''\rightarrow V''-m^{2}$)
evaluated at coincident points are given by
\begin{align}
a_{0} & =1,\nonumber \\
a_{2} & =(-V''[\phi_{c}]+m^{2}),\nonumber \\
a_{4} & =\left(-\frac{1}{6}\nabla^{2}V^{''}[\phi_{c}]+\frac{1}{2}\left(V''[\phi_{c}]-m^{2}\right)^{2}\right),\nonumber \\
a_{6} & =\frac{1}{7!}\left(-6\nabla^{2}\nabla^{2}V''+60\left(V''-m^{2}\right)\nabla^{2}V''+30\nabla^{\mu}V''\nabla_{\mu}V''+60\left(V''-m^{2}\right)^{3}\right).\label{eq:aScalar}
\end{align}
To evaluate the one loop correction to the 1PI action one needs the
following integral:
\begin{equation}
\int_{1/\Lambda^{2}}^{\infty}\frac{d\tau}{16\pi^{2}}e^{-\tau m^{2}}\tau^{n-3}=\frac{1}{16\pi^{2}}\frac{1}{m^{2\left(n-2\right)}}\Gamma\left(n-2,\frac{m^{2}}{\Lambda^{2}}\right),\label{eq:Incompletegamma}
\end{equation}
where the function on the RHS is the incomplete gamma function $\Gamma(\alpha,x)=\int_{x}^{\infty}e^{-t}t^{\alpha-1}$.

Thus we get for the 1PI action for a scalar field to one-loop $\Gamma_{\Lambda}[\phi_{c}]=S[\phi_{c}]+\Gamma_{\Lambda}^{(1)}[\phi_{c}]$,
with
\begin{align}
\Gamma_{\Lambda}^{(1)}[\phi_{c}] & =-\frac{1}{32\pi^{2}}\int d^{4}x\Large[m^{4}\Gamma\left(-2,\frac{m^{2}}{\Lambda^{2}}\right)+m^{2}\Gamma\left(-1,\frac{m^{2}}{\Lambda^{2}}\right)(-V''[\phi_{c}]+m^{2})+\nonumber \\
 & \Gamma\left(0,\frac{m^{2}}{\Lambda^{2}}\right)\left(\frac{1}{2}\left(V''[\phi_{c}]-m^{2}\right)^{2}\right)+\nonumber \\
 & \Gamma\left(1,\frac{m^{2}}{\Lambda^{2}}\right)\frac{1}{7!}\left(60\left(V''-m^{2}\right)\nabla^{2}V''+30\nabla^{\mu}V''\nabla_{\mu}V''+60\left(V''-m^{2}\right)^{3}\right)\Large].\label{eq:oneloopG}
\end{align}
The incomplete gamma functions here are given by the following expresssions:

\begin{align*}
\Gamma\left(-2,x\right) & =\frac{1}{2}\left[E_{1}(x)+e^{-x}\left(\frac{1}{x^{2}}-\frac{1}{x}\right)\right],\\
\Gamma(-1,x) & =-\left[E_{1}(x)-e^{-x}\frac{1}{x}\right],\\
\Gamma(0,x)=E_{1}(x) & \equiv\int_{x}^{\infty}e^{-t}t^{-1}dt=-\gamma-{\rm ln}x-\sum_{n=1}^{\infty}\frac{(-1)^{n}x^{n}}{nn!},\,\,|{\rm arg\,x|<\pi,}\\
\Gamma\left(1,x)\right) & =e^{-x}.
\end{align*}
The action to one loop is then $\Gamma(\phi_{c})=S(\phi_{c})+\Gamma_{\Lambda}^{(1)}[\phi_{c}]$.
The one loop quantum corrected equation of motion is 
\begin{equation}
\frac{\delta\Gamma}{\delta\phi_{c}}=\frac{\delta S}{\delta\phi_{c}}+\frac{\delta\Gamma_{\Lambda}^{(1)}}{\delta\phi_{c}}+O(\hbar^{2})=0.\label{eq:eofmtooneloop}
\end{equation}
It is important to note here that $\phi_{c}\ne<0|\hat{\phi}|0>$ which
is what one would have concluded if the initial and final states in
the path integral defining $\Gamma$ were both the vacuum state (with
$j=0$). Note that \eqref{eq:oneloopG} gives corrections to the kinetic
terms. It is instead the expectation in an arbitrary (non-Poincare
invariant) state just as the classical field in the classical action
is in general an arbitrary point in field space and not necessarily
the minimum of the potential. In other words it contains quantum corrected
dynamics and not just the quantum corrected potential minimum.

\subsection{A Euclidean digression: the definition of $\Gamma$ and convexity }

Consider a scalar field theory in Euclidean space. The generator of
connected correlation functions is 
\begin{equation}
e^{W_{OI}[j]}=<O|e^{j\cdot\hat{\phi}|_{-T/2}^{T/2}}|I>=\int[d\phi]\psi_{O}^{*}[\phi(+T/2)]\psi_{I}[\phi(-T/2)]e^{-S_{-T/2}^{T/2}}e^{j\cdot\phi|_{-T/2}^{T/2}}\label{eq:WOI}
\end{equation}
In the above $S_{-T/2}^{T/2}=\int_{-T/2}^{T/2}d\tau L[\phi]$ where
$L$ is the Lagrangian and $j.\phi|_{-T/2}^{T/2}=\int_{-T/2}^{T/2}d\tau\int d^{3}xj\phi$.
Also 
\[
\psi_{I}[\phi(-T/2)]=<\phi(-T/2)|I>,\,\,\psi_{O}[\phi(T/2)]=<\phi(T/2)|O>,
\]
 where the state $|\phi(\tau)>$ is an eigenstate of the field operator
$\hat{\phi}$ at Euclidean time $\tau$. Let us define the measure
in the functional integral as 
\begin{equation}
d\mu[\phi]=[d\phi]\psi_{O}^{*}[\phi(+T/2)]\psi_{I}[\phi(-T/2)]e^{-S_{-T/2}^{T/2}}\label{eq:dmu}
\end{equation}
with $[d\phi]$ defined by the usual time slicing in the interval
$[-T/2,T/2]$. The argument for convexity assumes a positive measure.
But this measure is not positive unless we take $\psi_{I}[\phi(-T/2)]=\psi_{O}[\phi(+T/2)].$
Of course this would be the case in equilibrium statistical mechanics
where the (Gibbs) partition function (with chemical potential $j$)
is calculated, and one evaluates a trace which implies periodic boundary
conditions in the functional integral. Euclidean QFT is then regarded
as being equivalent to the partition function at inverse temperature
$\beta$ (which in our case is $T$) being taken to infinity - i.e.
as the zero temperature limit. But then one loses the dynamics of
the system and is forced to the (global) minimum of the potential. 

In the general case the measure $d\mu$ is not positive and hence
the argument for convexity \citep{Iliopoulos:1974ur} based on Hoelder's
inequality is not valid. In particular 
\begin{equation}
\frac{\delta^{2}W_{OI}[j]}{\delta j(x_{1})\delta j(x_{2})}=<O|(\hat{\phi}(x_{1})-<O|\hat{\phi}(x_{1})|I>)(\hat{\phi}(x_{2})-<O|\hat{\phi}(x_{2})|I>)|I>_{j},\label{eq:d2W}
\end{equation}
may in fact be positive negative or zero as a matrix unless the state
$|O>$ is identified with the state $|I>$ which would be the case
if say as $T\rightarrow\infty$ (with $j$ being adiabatically turned
on and off) the system relaxes to the (unique) vacuum $|0>$ in the
(Euclidean) far past and far future. 
\[
\lim_{T\rightarrow\infty}|O,T/2>=\lim_{T\rightarrow\infty}|I,-T/2>=|0>,
\]
In the latter case of course the eigenvalues of the matrix \eqref{eq:d2W}
are positive definite and the Legendre transform to the quantum effective
action 
\begin{equation}
\Gamma[\phi_{c}]=-W[j]+j\cdot\phi_{c},\,\,\phi_{c}=\frac{\delta W[j]}{\delta j}=<\hat{\phi}>\Rightarrow\frac{\delta\Gamma}{\delta\phi_{c}}=j,\label{eq:Gamma}
\end{equation}
is well defined everywhere in field space.

However in this case the system is at the global minimum of the effective
potential and (once $j\rightarrow0)$) only the equilibrium configurations
can be discussed. In particular since the vacuum (in flat space field
theory) is Poincare invariant, the defining equation for $\phi_{c}$
in \eqref{eq:Gamma} shows that $\phi_{c}$ is space-time independent
once the external source is set to zero. Hence the last equation in
\eqref{eq:Gamma} is just an equation that fixes $\phi_{c}$ at the
minimum of the effective potential. 

In general though the effective action $\Gamma$ should contain the
full quantum dynamics of the theory - even in the absence of the external
source $j$ just as the classical action $S[\phi],$ (which after
all is $\lim_{\hbar\rightarrow0}\Gamma[\phi_{c}]$). In other words
the equation of motion
\[
0=\frac{\delta\Gamma}{\delta\phi_{c}}=\frac{\delta S}{\delta\phi_{c}}+O(\hbar)
\]
must give the full content of quantum dynamics - including tunneling
events. Indeed what we showed at the end of the last section is how
perturbative quantum dynamics arises for a scalar field theory. However
this obviously depends on the interpretation of $\phi_{c}$. In particular
one cannot as is usually done take $\phi_{c}=<0|\hat{\phi}|0>$ i.e.
the vacuum expectation value in a Poincare invariant vacuum for this
would just give us the equilibrium situation.

To get dynamics one should not identify the states (in the Heisenberg
picture) $|I>,|O>$ as ground states of the Hamiltonian (or indeed
as any energy eigenstate) since obviously the matrix element of a
field operator and hence the ``classical'' field $\phi_{c}$ are
either time-independent or have a trivial time dependence. Keeping
these states to be general states\footnote{The general definition of the quantum effective action is given for
example in chapter 23 of \citep{DeWitt:2003pm}. See also \citep{Parker:2009uva}.} (corrsponding for example to some field configuration at $t=\pm\infty$
) we would not necessarily have convexity, as pointed out above .
All quantities in the quantum effective action now depend on these
states so we have
\begin{equation}
\frac{\delta W_{OI}[j]}{\delta j}=<O|\hat{\phi}|I>_{j}\equiv\phi_{c}^{(OI)},\label{eq:phiOI}
\end{equation}
and 
\begin{equation}
\Gamma_{OI}[\phi_{c}^{(OI)}]=-W_{OI}[j]+j\cdot\phi_{c}^{(OI)}\Rightarrow\frac{\delta\Gamma_{OI}}{\delta\phi_{c}^{OI}}=j\label{eq:GammaOI}
\end{equation}
Note that even if $\hat{\phi}$ is Hermitian $\phi_{c}^{(OI)}$ will
in general be complex unless $|O>=|I>$.

However now the problem is that the Legendre transformation breaks
down whenever the Jacobian of the transformation becomes singular
i.e. when 
\begin{equation}
\frac{\delta\phi_{c}^{(OI)}(x_{1})}{\delta j(x_{2})}=\frac{\delta^{2}W_{OI}[j]}{\delta j(x_{2})\delta j(x_{1})}.\label{eq:Jacobian}
\end{equation}
has a zero eigen value. Indeed as we saw in eqn.\eqref{eq:d2W} the
RHS can have zero eigenvalues so that at these points the transformation
will not be well-defined. Nevertheless one can combine the definitions
\eqref{eq:GammaOI}\eqref{eq:WOI} to get a self-consistent definition
of $\Gamma$\citep{DeWitt:1964mxt,DeWitt:2003pm},
\begin{equation}
e^{-\Gamma_{OI}[\phi_{c}^{(OI)}]}=\int[d\phi]\psi_{O}^{*}[\phi(+\infty)]\psi_{I}[\phi(-\infty)]e^{-S[\phi]+\frac{\delta\Gamma_{OI}}{\delta\phi_{c}^{OI}}\cdot\phi}.\label{eq:GammaOI-1}
\end{equation}
At points where the jacobian is well defined this will coincide with
the usual definition and hence there we will have the connection to
the operator formalism. 

The initial and final state wave functions in the functional integral
cannot in general be replaced by free field ground state wave functions
as in Weinberg \citep{Weinberg:1995mt}. We may slightly generalize
this argument as follows. Let us write
\begin{equation}
\psi[\phi(\pm\infty)]=\exp\left\{ -\int d^{3}x\int d^{3}y\phi({\bf x},\pm\infty)K({\bf x},{\bf y})\phi({\bf y,\pm\infty)}+O(\phi^{4})\right\} \equiv e^{-\Omega(\pm\infty)},\label{eq:psipm}
\end{equation}
where $\Omega$ should be such that the wave function is normlizable.
If we assume that the interaction is to be ignored in the far future
and past (as is usual in definitions of the S-matrix) only the quadratic
term in $\Omega$ will be present. However we do not need to make
this assumption. IHowever one may still write \citep{Weinberg:1995mt}
\begin{equation}
\psi_{O}^{*}[\phi(+\infty)]\psi_{I}[\phi(-\infty)]=e^{-\left\{ \Omega(\infty)+\Omega(-\infty)\right\} }=\lim_{\epsilon\rightarrow+0}\exp\left[\epsilon\int_{-\infty}^{\infty}d\tau e^{-\epsilon|\tau|}\Omega(\tau)\right].\label{eq:iepsilon}
\end{equation}
In the Lorentzian formulation of QFT the wave function factor at the
two ends of the time line could be thought of as an $i\epsilon$ prescription
for doing perturbation theory. Note that in order to derive perturbation
theory one would simply have to replace the terms higher than quadratic
in the wave functions \eqref{eq:psipm} by the appropriate functional
derivatives with respect to $j$ as in \eqref{eq:ExpGamma_expansion}.

The main point of the above discussion was to show that unless one
imposes periodic (in time) boundary conditions the argument for convexity
does not hold. In quantum field theory one is not obliged to identify
the construction of the effective action $\Gamma$ in analogy with
equilibrium statistical mechanics at zero temperature. Of course the
price one has to pay is that the Legendre transformation is not well
defined everywhere in the space of classical sources $j$ and their
duals $\phi_{C}$, though the points at which it breaks down is a
set of measure zero.

\subsection{Curved space\label{subsec:Curved-space}}

Let us extend the above to curved space.

The basic formula for the quantum effective action will still be given
by \eqref{eq:ExpGamma_expansion} or \eqref{eq:Gammaexpansion} except
that $\eta_{\mu\nu}\rightarrow g_{c\mu\nu}$ the metric on space-time,
and the derivative operator $\nabla$ will be covariant with gauge
fields as well as the gravitational field. We will just focus in this
subsection on generalizing the regularized definition of the propagator
and the one-loop effective action to curved space. The expression
\eqref{eq:Kinverse1} is then replaced by\citep{DeWitt:1964mxt}\citep{DeWitt:2003pm},
\begin{equation}
{\bf K}^{-1}\left[\phi_{c};x,y\right]=-\int_{0}^{\infty}\frac{ds}{16\pi^{2}s^{2}}e^{-ism^{2}(\phi_{c}){\bf I}}e^{\frac{i}{2s}\sigma\left(x,y\right)}{\bf F}(x,y;is).\label{eq:Propagatorcurved}
\end{equation}
Here the so-called world function $\sigma(x,y)$ is half the squared
geodesic distance betwee the points $x,y$ - so in flat space it is
equal to $(x-y)^{2}/2$. The (matrix valued) function ${\bf F}$ is
not known in closed form but has an asymptotic expansion for short
proper time $is$ 
\begin{equation}
{\bf F}(x,y;is)\approx\sum_{n=0}^{\infty}{\bf a}_{2n}(x,y)\left(is\right)^{n}\approx{\bf a}_{0}(x,y)+{\bf a}_{2}(x,y)is+{\bf a}_{4}(x,y)\left(is\right)^{2}+\ldots.\label{eq:heatexpansion}
\end{equation}
The coefficients may be determined recursively by using the heat equation
\eqref{eq:Heateqn}\footnote{See for example \citep{DeWitt:1964mxt}. More efficient methods have
been developed in both the mathematics and the physics literature.
For a review of the heat kernel expansion and for references to the
literature see \citep{Vassilevich:2003xt}. This paper however only
considers the coincidence limits of the coefficients $a_{n}$ . For
a recent evaluation of the $a$'s at non-concident points even going
beyond the quadratic minimal diffential operators that typically occur
in two derivative theories see \citep{Barvinsky:2021ijq}.}. As before we have chosen to factor out a constant unit matrix proportional
to the lowest (squared) mass in the theory in order to provide us
with an infra-red regulator. As in the flat space case we can analytically
continue this integral to the imaginary half-line for space-like separated
points and then define it for all points including light-like separations
or coincident points by introducing a UV cutoff, so that we have for
the regulated propagator in curved space,
\begin{equation}
{\bf K}_{\Lambda}^{-1}\left[\phi_{c};x,y\right]=-i\int_{1/\Lambda^{2}}^{\infty}\frac{d\tau}{16\pi^{2}\tau^{2}}e^{-\tau m^{2}(\phi_{c}){\bf I}}e^{-\frac{1}{2\tau}\sigma\left(x,y\right)}{\bf F}(x,y;\tau).\label{eq:PropagatorcurvedRegulated}
\end{equation}
We can now write down the regulated expression for the 1-loop effective
action in curved space, the analog of the flat space expression \eqref{eq:Gamma1Reg}
and indeed it's given by the same expression as there, except that
coefficients $a_{2n}(x,x)$ would be more complicated - for instance
they will involve tangent bundle as well as gauge bundle curvatures.

\section{Vilkovisky DeWitt (VDW) Effective Action}

The definition of $\Gamma$ given in \eqref{eq:PIforGamma} is not
field redefinition independent, unlike the classical action which
is a scalar under general coordinate transformations (GCT) on field
space. This is because the last term in the exponent on the RHS is
not invariant since $\phi$ is a coordinate and not a vector. This
was remedied by Vilkovisky and a technical issue with the original
definition was corrected by DeWitt\footnote{For a recent clear discussion and references to the original literature
see \citep{Parker:2009uva}.}. The basic idea is to replace $\phi-\phi_{c}$ the coordinate difference
in field space by the tangent vector to the geodesic from $\phi_{c}$
to $\phi$ at $\phi_{c}$. This is just the derivative of the world
function in field space $\sigma^{i}\equiv\nabla^{i}\sigma(\phi_{c}.\phi)$
where $\sigma(\phi_{c}.\phi)$ is defined as in the previous section
but in field space rather than in space-time. In other words it is
half the geodesic distance between the two points $\phi_{c}$ and
$\phi$ and is a biscalar function of the two arguments. The derivative
is with respect to the first argument and the index was raised using
the background field space metric $G_{ij}[\phi_{c}]$. In the flat
limit we get $\sigma^{i}\left(\phi_{c},\phi\right)\rightarrow\left(\phi_{c}-\phi\right)^{i}$. 

The definition \eqref{eq:PIforGamma} is then replaced by 
\begin{equation}
e^{i\Gamma[\phi_{c}]}=\int[d\phi]e^{iS[\phi]-ij_{i}[\phi_{c}]\sigma^{i}\left(\phi_{c},\phi\right)}|_{j=-\delta\Gamma/\delta\phi_{c}}.\label{eq:VilkDeWaction}
\end{equation}
The last term in the exponent is more explicitly $\int d^{4}x\sqrt{g}j_{i}[\phi_{c}(x)]\sigma^{i}(\phi_{c}\left(x\right),\phi\left(x\right))$
and is invariant under both GCT of $\phi_{c}$ as well as of $\phi$
since $\sigma^{a}$ is a vector under GCT of the coordinates $\phi_{c}$
whilst it is a scalar under GCT of $\phi$, and $j[\phi_{c}]=-\delta\Gamma/\delta\phi_{c}$
is a covector under GCT of $\phi_{c}$. Also covariant derivatives/connections
on field space are defined to be compatible with the field space metric
$G_{ij}$.

\subsection{Gauge and gauge fixing independence}

Let us now discuss the gauge independece of $\Gamma$. The functional
integral in a gauge theory needs to be defined by the DeWitt-Faddeev-Popov
procedure. The gauge needs to be fixed and the compensating ghost
action needs to be added. This action is invariant under the BRST
transormation ${\cal S}\phi^{i}=R^{i}[\phi]$ where the index range
of $i$ has been extended to include the ghost fields and an auxiliary
field. ${\cal S}$ is a (fermionic) differential operator on field
space which is nilpotent i.e. ${\cal S}^{2}=0$. The total action
is then 
\begin{equation}
S[\phi]=S_{{\rm classical}}+{\cal S}\Psi[\phi]=S_{{\rm classical}}+\Psi[\phi]_{,i}R^{i}[\phi].\label{eq:Stotal}
\end{equation}
Here $\Psi$ is a fermionic functional of all the fields. The gauge
invariance of the original (``classical'') action and the nilpotency
of ${\cal S}$ implies the BRST invariance of the total action i.e.
${\cal S}S=S_{i}[\phi]R^{i}[\phi]=0$. In particular the metric on
the space of fields 
\begin{equation}
||\delta\phi||^{2}=\int\sqrt{g}d^{4}xG_{ij}[\phi]\delta\phi^{i}(x)\delta\phi^{j}(x)\label{eq:metric}
\end{equation}
 (and hence the measure in the functional integral) should be invariant.
This implies that $R^{i}$ is a Killing vector on field space, 
\begin{equation}
R_{i;j}+R_{j;i}=0,\Rightarrow R^{i}\,\,_{;i}=0.\label{eq:Rij}
\end{equation}

Changing the gauge fixing condition is tantamount to changing $\Psi$.
Consider the effect of such an (infinitesimal) change on $\Gamma$.
\begin{align}
e^{i\Gamma}i\delta_{\Psi}\Gamma & =\int[d\phi]\delta_{\Psi}e^{i\left\{ S\left[\phi\right]+\frac{\delta\Gamma}{\delta\phi_{c}^{i}}.\sigma^{i}\left(\phi_{c},\phi\right)\right\} }\nonumber \\
 & =\int[d\phi]\left\{ \delta\Psi_{i}[\phi]R^{i}[\phi]+\delta_{\Psi}\Gamma_{i}[\phi_{c}]\sigma^{i}\left(\phi_{c},\phi\right)\right\} e^{i\left\{ S\left[\phi\right]+\frac{\delta\Gamma}{\delta\phi_{c}^{i}}.\sigma^{i}\left(\phi_{c},\phi\right)\right\} }.\label{eq:deltaPsieiGamma}
\end{align}
Let us define $<A[\phi]>\equiv e^{-i\Gamma[\phi]}\int[d\phi]A[\phi]e^{i\left\{ S\left[\phi\right]+\frac{\delta\Gamma}{\delta\phi_{c}^{i}}.\sigma^{i}\left(\phi_{c},\phi\right)\right\} }$.
Also the analog of the usual relation $<\phi>=\phi_{c}$ is now\footnote{See for example Parker and Toms \citep{Parker:2009uva} eqn. 6.137.
This follows from eqn. 6.118 when we put $\phi^{*}=\bar{\phi}$. There
is a complication in the case of the DeWitt action but the above is
still expected to be true at least in perturbation theory - see chapter
7 section 7.7.2 of the above reference.} $<\sigma^{i}\left(\phi_{c},\phi\right)>=0$. Thus we have the following
relation 
\begin{equation}
\delta_{\Psi}\Gamma=<\delta\Psi_{,i}[\phi]R^{i}[\phi]>.\label{eq:deltaPsiGamma}
\end{equation}
Now we use integration by parts in the functional integral to shift
the derivative in the above from $\delta\Psi.$
\begin{align}
0 & =\int[d\phi]\nabla_{j}\left(\delta\Psi[\phi]R^{j}[\phi]e^{i\left\{ S\left[\phi\right]+\frac{\delta\Gamma}{\delta\phi_{c}^{i}}.\sigma^{i}\left(\phi_{c},\phi\right)\right\} }\right)\nonumber \\
 & =\int[d\phi]\left(\delta\Psi_{,j}[\phi]R^{j}[\phi]+\delta\Psi[\phi]R_{\,\,;j}^{j}[\phi]+\delta\Psi R^{j}[\phi]\frac{\delta\Gamma}{\delta\phi_{c}^{i}}.\frac{\delta\sigma^{i}\left(\phi_{c},\phi\right)}{\delta\phi^{j}}\right)e^{i\left\{ S\left[\phi\right]+\frac{\delta\Gamma}{\delta\phi_{c}^{i}}.\sigma^{i}\left(\phi_{c},\phi\right)\right\} }.\label{eq:partial}
\end{align}
In the above we used the gague invariance of the classical action.
Using \eqref{eq:Rij} (which implies that the second term above is
zero) and \eqref{eq:deltaPsiGamma} we then have for the gauge fermion
variation of the quantum effective action the relation.
\begin{equation}
\delta_{\Psi}\Gamma=-\frac{\delta\Gamma}{\delta\phi_{c}^{i}}<\delta\Psi R^{j}[\phi]\frac{\delta\sigma^{i}\left(\phi_{c},\phi\right)}{\delta\phi^{j}}>.\label{eq:deltaPsiGamma1}
\end{equation}
With the usual definition of $\Gamma$ the third factor inside $<>$
is $\delta_{j}^{i}$ and the relation becomes $\delta_{\Psi}\Gamma=-\frac{\delta\Gamma}{\delta\phi_{c}^{j}}<\delta\Psi R^{j}[\phi]>.$
However from this realtion all we can say for the usual definition
of $\Gamma$ is that it is gauge fixing independent at its extremum
$\delta\Gamma/\delta\phi_{c}=0$.

For the Vilkovisky-DeWitt action however we have\footnote{See Parker and Toms \citep{Parker:2009uva} eqn. 7.181.}
the relation 
\begin{equation}
R^{j}[\phi]\frac{\delta\sigma^{i}\left(\phi_{c},\phi\right)}{\delta\phi^{j}}\propto R^{i}[\phi_{c}]\label{eq:Vilkovisky}
\end{equation}
 Hence we get using the gauge invariance of $\Gamma[\phi_{c}]$ (assuming
background invariant gauge fixing),
\begin{equation}
\delta_{\Psi}\Gamma\propto-\frac{\delta\Gamma}{\delta\phi_{c}^{i}}R^{i}[\phi_{c}]=0,\label{eq:deltaPsiGamma2}
\end{equation}
showing the gauge fixing independence of $\Gamma.$ 

The DeWitt version of $\Gamma$ is actually somewhat more complicated
and is the price we have to pay if the metric on field space is non-trivial
(i.e. with non-zero Riemann tensor). In this case the last term in
the exponent of \eqref{eq:VilkDeWaction} is replaced by 
\begin{equation}
\frac{\delta\Gamma}{\delta\phi_{c}^{k}}C_{\,\,i}^{k}[\phi_{c}]\sigma^{i}\left(\phi_{c},\phi\right),\,\,C_{\,\,i}^{k}[\phi_{c}]=<\sigma^{k}\,\,_{;i}(\phi_{c},\phi)>.\label{eq:DeWittC}
\end{equation}
 The only change is in the last term of \eqref{eq:partial} and hence
eqn \eqref{eq:deltaPsiGamma1} becomes
\[
\delta_{\Psi}\Gamma=-\frac{\delta\Gamma}{\delta\phi_{c}^{k}}C^{k}\,\,\,_{i}[\phi_{c}]<\delta\Psi R^{j}[\phi]\frac{\delta\sigma^{i}\left(\phi_{c},\phi\right)}{\delta\phi^{j}}>\propto-\frac{\delta\Gamma}{\delta\phi_{c}^{k}}R^{i}[\phi_{c}]<\sigma^{k}\,\,_{;i}(\phi_{c},\phi)>.
\]
But we also have the relation\footnote{See Parker and Toms \citep{Parker:2009uva} eqn. 7.186.}
$R^{i}[\phi_{c}]\sigma^{k}\,\,_{;i}(\phi_{c},\phi)\propto R^{k}[\phi_{c}]$
so that we get 
\begin{equation}
\delta_{\Psi}\Gamma\propto-\frac{\delta\Gamma}{\delta\phi_{c}^{k}}R^{i}[\phi_{c}]<\sigma^{k}\,\,_{;i}(\phi_{c},\phi)>\propto-\frac{\delta\Gamma}{\delta\phi_{c}^{k}}R^{k}[\phi_{c}]=0,\label{eq:deltaPsiGamma3}
\end{equation}
thus establishing the gauge fixing independence of the DeWitt quantum
effective action.

\subsection{The formal perturbative expansion of $\Gamma$}

This sub-section is a summary of the discussion in section 6.6 of
Parker and Toms \citep{Parker:2009uva}. To construct the perturbation
theory of the VDW effective action we start with the following definition
(which generalizes \eqref{eq:VilkDeWaction}),
\begin{equation}
e^{i\Gamma[\phi^{*},\phi_{c}]}=\int d\mu\left[\phi^{*};\phi\right]e^{i\left\{ S\left[\phi\right]+\frac{\delta\Gamma}{\delta v^{i}}(v^{i}-\sigma^{i}\right\} }\label{eq:GeneralGamma}
\end{equation}
where 
\begin{align*}
\sigma^{i} & \equiv\sigma^{i}\left[\phi^{*};\phi\right],\\
v^{i} & \equiv\sigma^{i}\left[\phi^{*};\phi_{c}\right].
\end{align*}
The measure in the functional integral is
\[
d\mu\left[\phi^{*};\phi\right]=\prod_{i}d\sigma^{i}|g(\phi^{*})|^{1/2}=\prod_{i}d\phi^{i}|g(\phi)|^{1/2}\Delta\left[\phi^{*};\phi\right].
\]
and $\Delta\left[\phi^{*};\phi\right]=|g(\phi^{*})|^{-1/2}|g(\phi)|^{-1/2}{\rm S}\det\left(-\frac{\delta^{2}\sigma\left[\phi^{*};\phi\right]}{\delta\phi^{*}\delta\phi}\right)$
is the Van Vleck-Morette determinant. The DeWitt version of the effective
action is obtained by putting $\phi^{*}=\phi_{c}$ in \eqref{eq:GeneralGamma}.
However it is more convenient to use the general form above to construct
the perturbation series \citep{Burgess:1987zi,Rebhan:1987cd}. Also
the fields now range over the original physical fields as well as
the DeWitt-Faddeev-Popov ghosts and auxiliary fields that one has
in the BRST formalism. 

First we take the action to be a functional of $\phi^{*}$ and $\sigma^{i}$.
i.e.
\begin{align}
S[\phi] & =\hat{S}[\phi^{*};\sigma^{i}\left[\phi^{*};\phi\right]]=\sum_{n=0}^{\infty}\frac{\left(-1\right)^{n}}{n!}S_{;\left(i_{1}\ldots i_{n}\right)}[\phi^{*}]\sigma^{i_{1}}\left[\phi^{*};\phi\right]\ldots\sigma^{i_{n}}\left[\phi^{*};\phi\right]\label{eq:covariantTaylor}\\
 & =\hat{S}\left[\phi^{*};v^{i}\right]+\sum_{n=1}^{\infty}\frac{1}{n!}\frac{\delta^{n}\hat{S}\left[\phi^{*};v^{i}\right]}{\delta v^{i_{1}}\ldots\delta v^{i_{n}}}\left(\sigma^{i_{1}}-v^{i_{1}}\right)\ldots\left(\sigma^{i_{n}}-v^{i_{n}}\right),\label{eq:Taylor}
\end{align}
where the first relation is a covariant Taylor expansion around $\sigma^{i}=0$
($S_{;\left(i_{1}\ldots i_{n}\right)}$ are symmetrized covariant
derivatives) and in the last relation we have expanded in a ordinary
Taylor series around $\sigma^{i}=v^{i}$. We also have the following
relations from the above,
\begin{align}
\hat{S}\left[\phi^{*};v^{i}\right] & =S\left[\phi_{c}\right]\label{eq:ShatSphic}\\
\frac{\delta^{n}\hat{S}\left[\phi^{*};v^{i}\right]}{\delta v^{i_{1}}\ldots\delta v^{i_{n}}}\mid_{v=0} & =\left(-1\right)^{n}S_{;\left(i_{1}\ldots i_{n}\right)}[\phi_{c}]\label{eq:Taylorcoeffs}
\end{align}

Let us now rewrite \eqref{eq:GeneralGamma} using the above relations.
\begin{equation}
e^{i\Gamma[\phi^{*},\phi_{c}]}=\int d\mu\left[\phi^{*};\phi\right]e^{iA\left[\phi_{c};\phi^{*};\sigma\right]},\label{eq:GeneralGamma2}
\end{equation}
 where 
\begin{align}
A\left[\phi_{c};\phi^{*};\sigma\right] & =S[\phi_{c}]+\frac{1}{2!}\hat{S}_{,ij}\left[\phi^{*};v^{i}\right]\sigma^{i}\sigma^{j}+\sum_{n=3}^{\infty}\frac{1}{n!}\hat{S}_{,i_{1}\ldots i_{n}}\left[\phi^{*};v^{i}\right]\sigma^{i_{1}}\ldots\sigma^{i_{n}}\nonumber \\
 & +\left(\hat{S}_{,i}\left[\phi^{*};v^{i}\right]-\hat{\Gamma}_{,i}\left[\phi^{*};v^{i}\right]\right)\sigma^{i}.\label{eq:A}
\end{align}
The derivatives in the above are with respect to $v$ keeping $\phi^{*}$
fixed and $\hat{\Gamma}\left[\phi^{*};v^{i}\right]=\Gamma\left[\phi^{*};\phi_{c}\right]$.
Using the same procedure as in the flat field space case we have
\begin{align}
e^{i\Gamma[\phi^{*},\phi_{c}]} & =e^{iS[\phi_{c}]}e^{i\sum_{n=3}^{\infty}\frac{1}{n!}\hat{S}_{,i_{1}\ldots i_{n}}\left[\phi^{*};v^{i}\right]\frac{\delta}{i\delta j^{i_{1}}}\ldots\frac{\delta}{i\delta j^{i_{n}}}}\int d\mu\left[\phi^{*};\phi\right]e^{i\left\{ \frac{1}{2!}\hat{S}_{,ij}\left[\phi^{*};v^{i}\right]\sigma^{i}\sigma^{j}+j_{i}\sigma^{i}\right\} }\vert_{j_{i}=\left(\hat{S}_{,i}\left[\phi^{*};v^{i}\right]-\hat{\Gamma}_{,i}\left[\phi^{*};v^{i}\right]\right)}\nonumber \\
 & =e^{iS[\phi_{c}]}e^{-\frac{1}{2}{\rm STr\ln\hat{S}^{,i}\,j\left[\phi^{*};v^{i}\right]}}e^{i\sum_{n=3}^{\infty}\frac{1}{n!}\hat{S}_{,i_{1}\ldots i_{n}}\left[\phi^{*};v^{i}\right]\frac{\delta}{i\delta j^{i_{1}}}\ldots\frac{\delta}{i\delta j^{i_{n}}}}e^{-\frac{i}{2}j_{i}\hat{\Delta}^{ij}j_{j}}\vert_{j_{i}=\left(\hat{S}_{,i}\left[\phi^{*};v^{i}\right]-\hat{\Gamma}_{,i}\left[\phi^{*};v^{i}\right]\right)}.\label{eq:GeneralGammaexpansion}
\end{align}
Here we have defined $\hat{S}^{,i}\,_{j}\left[\phi^{*};v^{i}\right]=g^{ik}[\phi^{*}]\hat{S}_{,kj}\left[\phi^{*};v^{i}\right]$
and $\hat{\Delta}$ is the inverse of the Kinetic operator $S_{,ij}\left[\phi^{*};v^{i}\right]$.

Now set $\phi^{*}=\phi_{c}$ and define the DWV effective action as
$\Gamma_{D}[\phi_{c}]=\Gamma\left[\phi^{*}=\phi_{c},\phi_{c}\right].$
Thus we get our final (unregularized) expression by setting $v=0$
in the above and using\eqref{eq:Taylorcoeffs},
\begin{equation}
e^{i\Gamma_{D}[\phi_{c}]}=e^{iS\left[\phi_{c}\right]}e^{-\frac{1}{2}{\rm STr\ln S^{;i}\,j\left[\phi_{c}\right]}}e^{i\sum_{n=3}^{\infty}\frac{1}{n!}S_{;i_{1}\ldots i_{n}}\left[\phi_{c}\right]\frac{\delta}{i\delta j^{i_{1}}}\ldots\frac{\delta}{i\delta j^{i_{n}}}}e^{-\frac{i}{2}j_{i}\Delta_{c}^{ij}j_{j}}\vert_{j_{i}=\left(S_{,i}\left[\phi_{c}\right]-\Gamma_{D,i}\left[\phi_{c}\right]\right)},\label{eq:GammaD}
\end{equation}
with 
\[
\Delta_{c}^{ij}S_{;jk}[\phi_{c}]=\delta_{l}^{i}.
\]
Note that $\Delta_{c}^{ij}$ is the $ij$ matrix element of the operator
$K^{-1}$ discussed in subsection \eqref{subsec:Curved-space}.

\subsection{Heat Kernel regularization of $\Gamma$}

We now use the same argument as before to define the regularized perturbation
expansion of the DeWitt-Vilkovisky effective action. Thus the UV heat
kernel regulated version of the perturbative expansion is 
\begin{equation}
e^{i\Gamma_{D}[\phi_{c}]}=e^{iS_{\Lambda}\left[\phi_{c}\right]}e^{-\frac{1}{2}{\rm Tr\ln S_{\Lambda}^{;i}\,_{j}\left[\phi_{c}\right]}}e^{i\sum_{n=3}^{\infty}\frac{1}{n!}S_{\Lambda;i_{1}\ldots i_{n}}\left[\phi_{c}\right]\frac{\delta}{i\delta j^{i_{1}}}\ldots\frac{\delta}{i\delta j^{i_{n}}}}e^{-\frac{i}{2}j_{i}\Delta_{\Lambda c}^{ij}j_{j}}\vert_{j_{i}=\left(S_{\Lambda,i}\left[\phi_{c}\right]-\Gamma_{D,i}\left[\phi_{c}\right]\right)},\label{eq:GammaRegulated}
\end{equation}
where the regularized propagator is given by \eqref{eq:PropagatorcurvedRegulated}.
Correspondingly the regularized expression for the one-loop effective
action (i.e. the exponent in the second factor above) is obtained
from (we've put ${\bf K}\equiv[S_{;ij}]$)
\[
\ln{\bf K}\left[\phi_{c};x,y\right]_{\Lambda}=i\int_{1/\Lambda^{2}}^{\infty}\frac{d\tau}{16\pi^{2}\tau^{3}}e^{-\tau m^{2}(\phi_{c}){\bf I}}e^{-\frac{1}{2\tau}\sigma\left(x,y\right)}{\bf F}(x,y;\tau).
\]
ie. 
\[
{\rm STr}\ln S^{;i}\,_{j}\left[\phi_{c}\right]_{\Lambda}=i\int\sqrt{|g|}d^{4}x\int_{1/\Lambda^{2}}^{\infty}\frac{d\tau}{16\pi^{2}\tau^{3}}{\rm tr}e^{-\tau m^{2}(\phi_{c}){\bf I}}{\bf F}(x,x;\tau).
\]
Note that as was the cae with \eqref{eq:PropagatorcurvedRegulated}
we've included an infra-red regulator mass term to define these integrals.
However in defining the Wilsonian effective action below we may dispense
of it. Note that we've imposed $\Lambda$ dependence on the classical
action (which should be regarded as an initial action at the RG time
$\ln\Lambda$) in order to ensure that $\Gamma$ is independent of
$\Lambda$. Requiring this independence is of course what keads the
RG eqns for the coupling constants.

\section{Wilsonian effective action from DWV action}

The Wilsonian effective action is obtained by integrating the modes
from some initial UV $\Lambda$ down to some lower scale say $\Lambda'<\Lambda.$
Thus we replace $\Gamma$ in \eqref{eq:GammaRegulated} by the action
at the scale $\Lambda'$. 
\begin{equation}
e^{iS_{\Lambda'}[\phi_{c}]}=e^{iS_{\Lambda}\left[\phi_{c}\right]}e^{-\frac{1}{2}{\rm {\rm STr}\ln S^{;i}\,_{j}\left[\phi_{c}\right]_{\Lambda}^{\Lambda'}}}e^{i\sum_{n=3}^{\infty}\frac{1}{n!}S_{\Lambda;i_{1}\ldots i_{n}}\left[\phi_{c}\right]\frac{\delta}{i\delta j^{i_{1}}}\ldots\frac{\delta}{i\delta j^{i_{n}}}}e^{-\frac{i}{2}j_{i}\Delta_{c}^{ij}\vert_{\Lambda}^{\Lambda'}j_{j}}\vert_{j_{i}=\left(S_{\Lambda,i}\left[\phi_{c}\right]-S{}_{\Lambda',i}\left[\phi_{c}\right]\right)}\label{eq:WEA}
\end{equation}
where the cutoff propagators and one loop determinant now have an
infrared cut off as well,
\begin{equation}
{\bf K}^{-1}\left[\phi_{c};x,y\right]_{\Lambda}^{\Lambda'}=i\int_{1/\Lambda^{2}}^{1/\Lambda'^{2}}\frac{d\tau}{16\pi^{2}\tau^{2}}e^{-\tau m^{2}(\phi_{c}){\bf I}}e^{-\frac{1}{2\tau}\sigma\left(x,y\right)}{\bf F}(x,y;\tau).\label{eq:K-1cutoff}
\end{equation}
\begin{equation}
\ln{\bf K}\left[\phi_{c};x,y\right]_{\Lambda}^{\Lambda'}=i\int_{1/\Lambda^{2}}^{1/\Lambda'^{2}}\frac{d\tau}{16\pi^{2}\tau^{3}}e^{-\frac{1}{2\tau}\sigma\left(x,y\right)}{\bf F}(x,y;\tau).\label{eq:lnKcutoff}
\end{equation}
ie. 
\begin{equation}
{\rm Tr}\ln S^{;i}\,_{j}\left[\phi_{c}\right]_{\Lambda}=i\int\sqrt{|g|}d^{4}x\int_{1/\Lambda^{2}}^{1/\Lambda'^{2}}\frac{d\tau}{16\pi^{2}\tau^{3}}{\rm tr}{\bf F}(x,x;\tau).\label{eq:TrlnKcutoff}
\end{equation}
Note that in the limit $\Lambda'\rightarrow\Lambda$ the above relations
and the definition\eqref{eq:WEA} are consistent with $S_{\Lambda'}\rightarrow S_{\Lambda}$.

\paragraph{Note added:}

While preparing these notes for posting, a paper \citep{Collison:2025pff}
which also discussed the guage and gauge fixing independence of the
Vilkovisky-DeWitt action (for flat space gauge theories) appeared
in ArXiv.

\bibliographystyle{apsrev}
\nocite{*}
\bibliography{myrefs}

\end{document}